\documentclass[11pt,a4paper]{article}

\usepackage[dvips]{graphics}

\newcommand{\ybox}[2]	{
 \begin{center}
 \resizebox{!}{#1\textheight}
{\includegraphics{#2.eps}}
 \end{center}		}

\begin{document}
\thispagestyle{empty}

\begin{center}

{\LARGE \bf The rate of cosmic ray showers at large zenith angles: a step towards the detection of 
ultra-high energy neutrinos by the Pierre Auger Observatory}

\end{center}

\begin{center}
{\bf M. Ave, R.A.~V\'azquez, and E.~Zas}\\
{\it Departamento de F\'\i sica de Part\'\i culas,\\
Universidad de Santiago, 15706 Santiago de Compostela, Spain\\}
{\bf J.A.~Hinton and A.A.~Watson}\\
{\it Department of Physics and Astronomy\\
 University of Leeds, Leeds LS2 9JT ,UK \\}
\end{center}

\begin{abstract}

It is anticipated that the Pierre Auger Observatory can be used to
detect cosmic neutrinos of $> 10^{19}$~eV that arrive at very large
zenith angles. However showers created by neutrino interactions close
to the detector must be picked out against a background of similar
events initiated by cosmic ray nuclei.  As a step towards
understanding this background, we have made the first detailed
analysis of air showers recorded at Haverah Park (an array which used 
similar detectors to those planned for the Auger Observatory) with zenith angles
above $60^{\circ}$.  We find that the differential shower rate from
$60^{\circ}$ to $80^{\circ}$ can be predicted accurately when we adopt
the known primary energy spectrum above $10^{17}$~eV and assume the
QGSJET model and proton primaries.  Details of the calculation are
given.
 
\end{abstract}

\vspace{1ex}

\section{Introduction}
\label{intro.sec}

Horizontal air showers have been studied for many years for several
different reasons \cite{Tokyo, EASTOP, CYGNUS}.  Much of the relevance
of horizontal air showers induced by cosmic rays is in the
understanding of the background against which high energy neutrino
showers could be detected. Although no neutrino events can be expected
in the Haverah Park data set, even for the most optimistic neutrino
flux predictions, it has recently been shown that relevant bounds
could be obtained with the Auger Observatories \cite{Auger,Capelle}.

Horizontal cosmic ray showers are of great interest in their own right
for two principle reasons. Firstly the acceptance of an air-shower
array could be doubled if events above $60^{\circ}$ can be adequately
analysed. Secondly because, uniquely, they deal with surviving
particles that are created very close to the shower core, they
complement the information obtained in the study of near vertical
cosmic ray showers. Moreover understanding the azimuthal asymmetries
at large zenith angles 
[6,7,8,9,10 (hereafter Paper I)]
can lead to significant improvement of ultra high energy shower
analysis at moderate zenith angles ($30^{\circ} - 60^{\circ}$)
\cite{Dedenko}.

The Haverah Park array, being made of 1.2~m deep water-\v Cerenkov
tanks \cite{haverah}, is quite possibly the array detector so far
constructed which is best suited on geometrical considerations for the
analysis of very large inclined showers. Moreover it can be considered
as an early prototype of the Auger Observatories which will employ
water-\v Cerenkov tanks of identical depth. The quantitative aspects
of our results are very specific to the water-\v Cerenkov technique.

During the 14 years of operation of the Haverah Park array
\cite{Lawrence} which are considered here nearly 10000 air showers
were recorded for $\theta > 60^{\circ}$.  The analysis of these data
was difficult because of the complex geomagnetic field effects which
distort the circular symmetry of air showers, the relatively small
size of the array (12~km$^{2}$) and the limited computing power then
available, both for data analysis and simulation. A single remarkable
event, perhaps of 10$^{20}$~eV, at $85^\circ$ from the zenith which
triggered 20 of the array detectors was extensively discussed
\cite{Hillas}, but no systematic study was made with those data for which
the initial analysis gave a zenith angle $\theta > 60^{\circ} $.

Here we present the results of the first systematic analysis of
horizontal showers from primaries of energy greater than $10^{17}$~eV
at Haverah Park with zenith angles exceeding $60^{\circ}$, comparing
them to those expected from the known cosmic ray spectrum. 
Preliminary results of this work were reported in \cite{utah99}. 
For the calculation of the
expected rates we make use of a parameterization for the muon number
densities at ground level described elsewhere (Paper I). We take
into account two possible compositions (proton and iron nuclei) and
use alternative models for the high energy interactions namely
QGSJET~\cite{qgsjet} and SIBYLL 1.6~\cite{sibyll}.  

The article is organized as follows: In section \ref{emag} we discuss
the relative signals and fluctuations expected from the
electromagnetic (electrons and photons) and muon components in
horizontal air showers (HAS) induced by hadrons and in section
\ref{model.sec} we address the muon component of HAS.  In
section~\ref{simulation} we discuss the detector simulation paying
particular attention to the origin of the different parts of the
signal from the individual tanks, stressing the differences between
vertical and horizontal particle signals. In section \ref{method} we
describe in detail the procedure implemented to obtain a prediction
for the horizontal shower rate and in section \ref{results} we give
the results obtained in the different cases.  Our conclusions are
presented in section \ref{conclusions}.

\section{Muon and Electromagnetic Components in Horizontal Air Showers}
\label{emag}

An air shower induced by a proton or a nucleus can be qualitatively
understood as a pion shower that is continuously feeding an
electromagnetic component (photons, electrons, and positrons) through
$\pi^0$ decay, and both a muonic and a neutrino component through
charged pion decay. At ground level the typical number density ratio
of photons to electrons in a vertical shower is stable. For a low
energy threshold of 90~keV this ratio has a value typically between
$20:1$ and $30:1$ for proton or iron as the distance to shower axis
$r$ changes from 100~m to 1~km. This value is characteristic of
electromagnetic showers. The ratio of electrons to muons on the other
hand depends strongly on $r$. For example for a $ 10^{19}~$eV proton
shower it drops from $\sim 100:1$ to $\sim 3:1$ as $r$ rises in the
same interval.

In spite of outnumbering the muon component, the   
average electron and photon energies are typically in the few MeV 
range, compared to GeV and above for muons. As a result the relative 
contributions to the signal in a water-\v Cerenkov tank are not so different 
because the water-\v Cerenkov technique is on average much more 
efficient at converting single muons into signals than electrons 
or photons. 

\begin{figure}
\ybox{0.3}{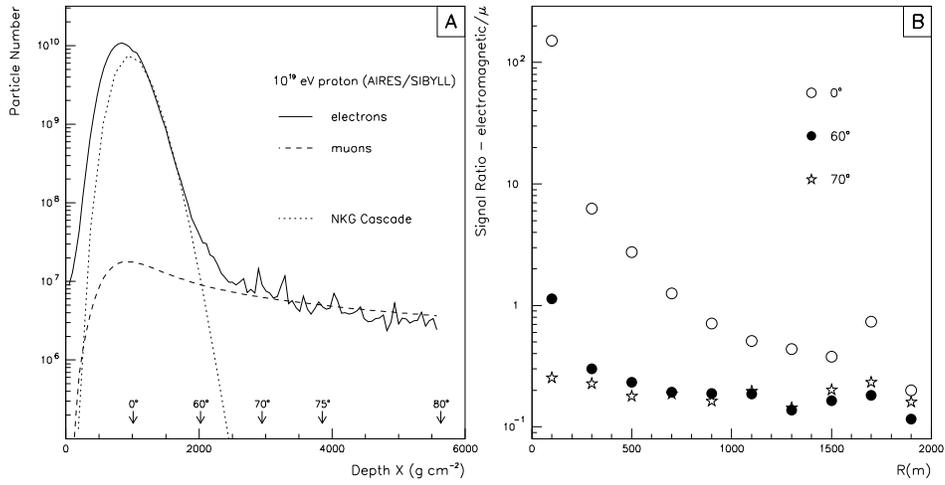}
\caption{ 
A) The longitudinal development of the muon and electron 
components averaged over 100, $10^{19}$~eV, showers. At depths 
exceeding 3000 
g cm$^{-2}$, or equivalently for zenith angles greater than 
70$^{\circ}$, the electromagnetic component is mainly due to 
muon decay. B) The ratio of the electromagnetic to muon 
contributions to water-\v Cerenkov signal as a function of distance 
from the shower axis (see section~4 for details of the water tank 
simulation). The large scatter in the points beyond 1500 m is 
statistical.}
\label{shower}
\end{figure}

Direct measurements with Fly's Eye~\cite{flyeseye} have shown that
above $10^{17}$~eV the shower maximum is usually between
600~g~cm$^{-2}$ and sea level ($\sim 1000$ g~cm$^{-2})$.  As the
zenith angle varies from the vertical, $\theta=0^\circ$, to the
horizontal, $\theta=90^\circ$, direction, the slant matter depth rises
to $\sim 36000$~g~cm$^{-2}$ and showers are observed well past
maximum. The behavior of the electromagnetic and muon components
beyond shower maximum is shown in Fig.~\ref{shower}A.  While the
electromagnetic component of an air shower becomes exponentially
attenuated with depth, the muons which do not decay propagate
practically unattenuated to the ground, except for energy loss and
geomagnetic deviations. As a result the ratio of the electromagnetic
to the muon component of an air shower drops as the zenith angle
increases up to $60^{\circ}$. Above this angle the ratio levels out
because the muons themselves produce electromagnetic particles.  The
remaining electromagnetic component is mainly due to muon decay and to
a smaller extent to hadronic interactions, pair production, and
bremsstrahlung.

We have studied the effects of muon decay through Monte Carlo
simulation using AIRES \cite{AIRES}. Indeed even at $60^\circ$ the
electromagnetic component of a $10^{19}$~eV proton shower which can be
directly associated to $\pi^0$ decay is already low and confined
within a relatively small region of about 200~m around shower axis. At
higher zenith angles this component can be neglected and the muon
decay contribution becomes stable, roughly at a proportion of 0.8
electrons for each muon, more than three times below the number
obtained for vertical showers 1~km away from the shower axis.  Unlike
the electromagnetic component from pion decay in
vertical showers, the lateral distribution follows that of the muons
rather closely. This is not difficult to understand as the
decay muons give rise to relatively small electromagnetic sub-showers
that preserve the muon spatial distribution. The energy distribution
of the electromagnetic component is essentially the same as that of an
electromagnetic air shower (i.e. for vertical showers) for the same
reason.

The contribution of muon interactions to the electromagnetic component
can be simply estimated by considering the muon energy spectrum of a
single shower and folding it analytically with the bremsstrahlung
cross section and the Greisen parameterization, see \cite{ZHV}. For an
$80^{\circ}$ zenith and $10^{19}~$eV proton shower the total number of
electrons and positrons ($N_e$) obtained is about $2.5~10^{4}$. These
are mostly due to the muons in the energy range between 30~GeV and
500~GeV. As this component arises also from electromagnetic sub-showers
its energy distribution should also reflect that of electromagnetic
cascades.  If we multiply, conservatively, the total number of electrons
by a factor three to account for the two other muon interactions, it
is still a factor of $\sim$50 below the number of muons in the shower
(see figure~\ref{shower}A).

The contribution to the signal from the electromagnetic component at
high zenith angles is in the end small and dominated by muon decay. It has
been simulated using WTANK \cite{Wtank} using the muon and
electromagnetic signals from different zenith angle showers generated
with AIRES \cite{AIRES} as an input. In Fig.~\ref{shower}B we show the
simulated ratio of electromagnetic to muon signal in a \v Cerenkov
tank of 1.2~m depth (as used in Haverah Park and planned for the 
Auger Observatory) as a function of
distance to the shower axis for a vertical shower compared to two
inclined showers. The shower particles have been fed through the
tank simulation as if they were coming from the vertical direction to
eliminate geometric tank effects. The results illustrate the genuine
decrease in the electromagnetic to muon signal ratio due only to the
shower composition varying with zenith angle. Even at distances from
the shower axis of order $\sim 1.5$~km the ratio is about a factor 3
smaller than for vertical showers.

As will be shown in section~\ref{simulation}, in the Haverah Park
detectors the relative electromagnetic signal drops even further as
the zenith angle increases because of geometric effects. At large
zenith angles the muon track length is enhanced compared with that in
the vertical direction and the output signal increases accordingly. In
the analysis that follows we will consider a track dependent, but
otherwise constant correction due to the electromagnetic component
from muon decay, in agreement with \cite{Dedenko}, and neglect that
from all muon interactions in the atmosphere.

\section{The muon component of horizontal showers}
\label{model.sec}

The behavior of the muonic component at high zenith angles becomes
extremely complex because of magnetic field effects. It is
described in detail in Paper I where an accurate model is
presented to account for the average muon number densities at ground
level. The inputs for this model are the lateral distribution function
(LDF), the average muon energy as a function of radius (${\cal E}
(r)$) and the mean distance to the muon production point, all
evaluated in the absence of magnetic effects.  The model is based on
an anticorrelation between the average muon energy and distance of
the muon from the core. We have generated these inputs, using AIRES, for two
primaries proton and iron, and two hadronic models Quark Gluon String
Model (QGSJET) and SIBYLL and for primary energies in the range
$10^{16}~$eV to $10^{20}$~eV.  The muon energy distribution at a fixed
distance to the shower axis is assumed to have a log-normal
distribution of width 0.4 (Paper I).

\begin{figure}
\ybox{0.5}{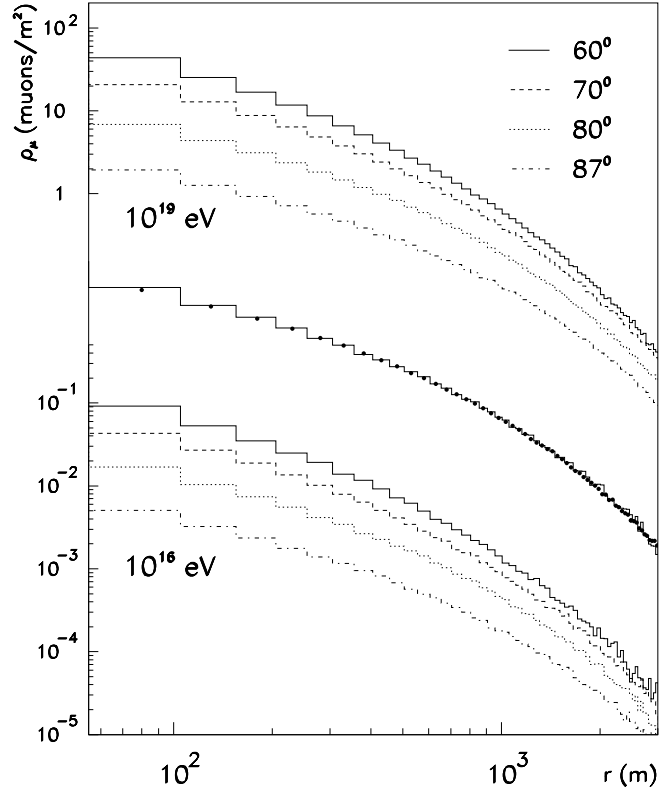}
\caption{Lateral distribution functions for primary protons of 
energy 10$^{16}$ eV compared with that for 10$^{19}$ eV protons 
for zenith angles of 60$^{\circ}$, 70$^{\circ}$, 80$^{\circ}$, and 
87$^{\circ}$ using SIBYLL. The two lateral distribution functions 
plotted in the centre of the figure are from 80$^\circ$ showers of 
10$^{16}$ eV (histogram) and 10$^{19}$ eV (dots) normalized 
using the relationship given in table~1 to emphasize the very close
similarity in shape.}
\label{LDF}
\end{figure}

\begin{figure}
\ybox{0.35}{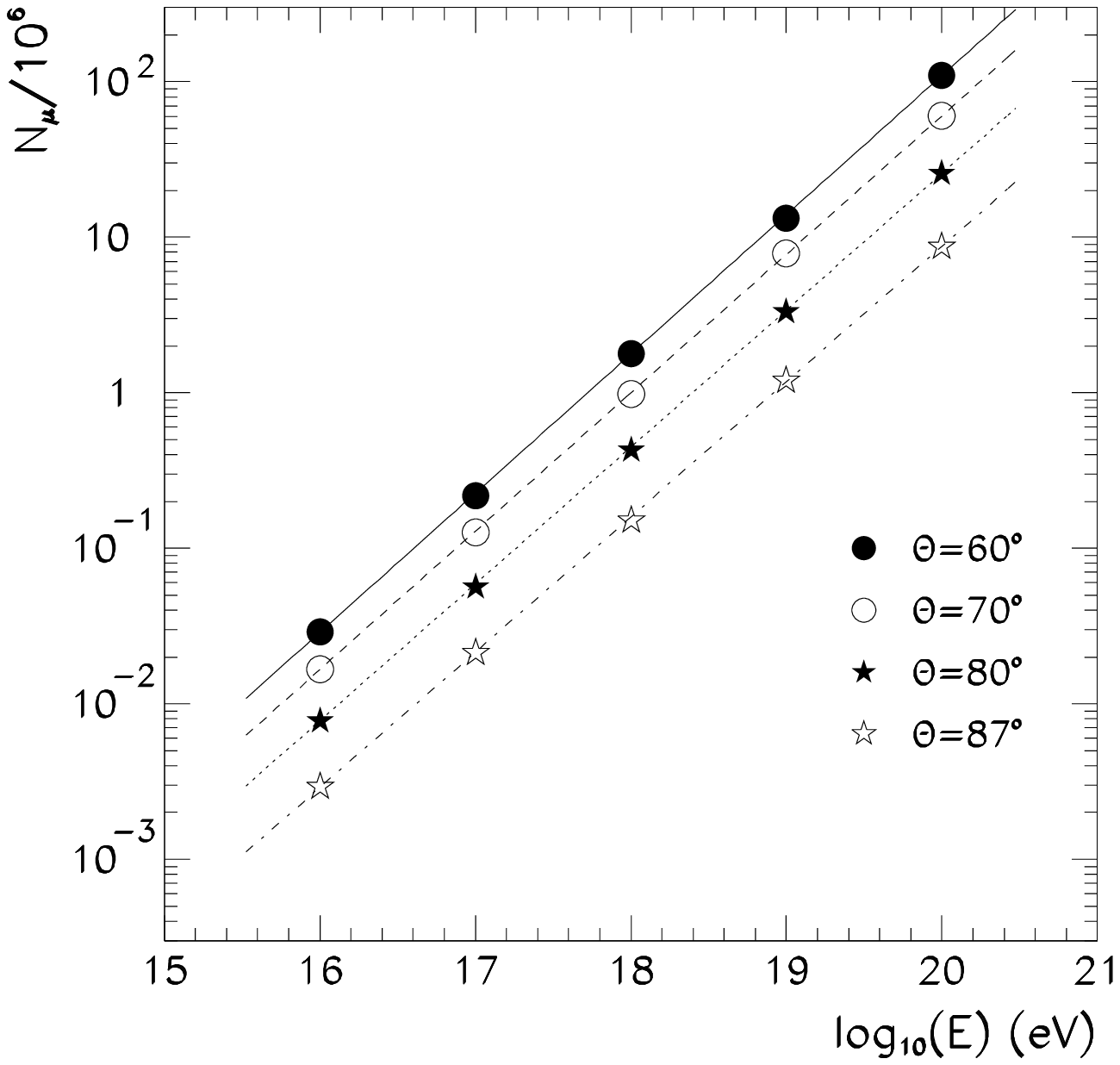}
\caption{The relationship of total muon number to primary energy 
for protons of four zenith angles using the SIBYLL model.}
\label{scaling}
\end{figure}

Both the ${\cal E} (r)$ and the shape of the LDF are essentially
invariant over this large energy interval and only mildly dependent on
zenith angle as shown in Fig.~\ref{LDF}.  As a result the dependence
on shower energy of the muon number density distributions with the
magnetic field can be parameterized with an absolute normalization to
a high level of precision. The dependence of the normalization factor
with energy can be obtained by monitoring the total muon number in the
showers. The results are plotted in Fig.~\ref{scaling} for four
different zenith angles. The normalization or energy scaling can be
taken into account by a relation of the following form:
\begin{equation}
N_\mu=N_0~E^{\beta}
\label{Escaling}
\end{equation}
where $N_0$ and $\beta$ are constants for a given model and mass 
composition as shown in Table~\ref{nmu.tab}. 

Fluctuations can enhance trigger rates for air showers produced by
lower energy primaries because of the steep cosmic ray spectrum
\cite{ParenteZas}. The fluctuations to high number of particles allow
the more numerous low energy showers to trigger. We have studied muon
number fluctuations at ground level and how they depend on shower
development (mean muon production height) and average muon energy. We
have found that the mean energy correlates strongly with production
height but that most of the number density fluctuations can be
accounted for by fluctuations in the total muon number. This is mainly
due to fluctuations in the neutral to charged pion ratios in the first
interactions. According to the results shown in Table~(1) of Paper I
it is sufficient to implement a $20 \%$ RMS fluctuation
in the average total muon number as obtained with Eq.~\ref{Escaling}.

Although the actual distributions obtained with
simulation do show a long tail to low muon numbers (up to a factor of
4 reduction), such showers are not expected to be relevant for
triggering the array precisely because they have so few muons. As a
result it is sufficient to assume a gaussian distribution.

\begin{table}
\begin{center}
\begin{tabular}{|lrcc|} \hline
Model & A & $\beta$ & $N_{\mu}$ ($10^{19}$ eV) \\\hline\hline
SIBYLL & 1 & 0.880 & 3.3 10$^{6}$ \\
       & 56 & 0.873 & 5.3 10$^{6}$ \\ \hline
QGSJET & 1 & 0.924 & 5.2 10$^{6}$ \\
       & 56 & 0.906 & 7.1 10$^{6}$ \\ \hline
\end{tabular}
\end{center}
\caption{Relationship between muon number and primary energy for 
different models and primary masses (see equation~\ref{Escaling}).}
\label{nmu.tab}
\end{table}

\section{Detector simulation}
\label{simulation}

\subsection{The Haverah Park Array}

\begin{figure}
\ybox{0.35}{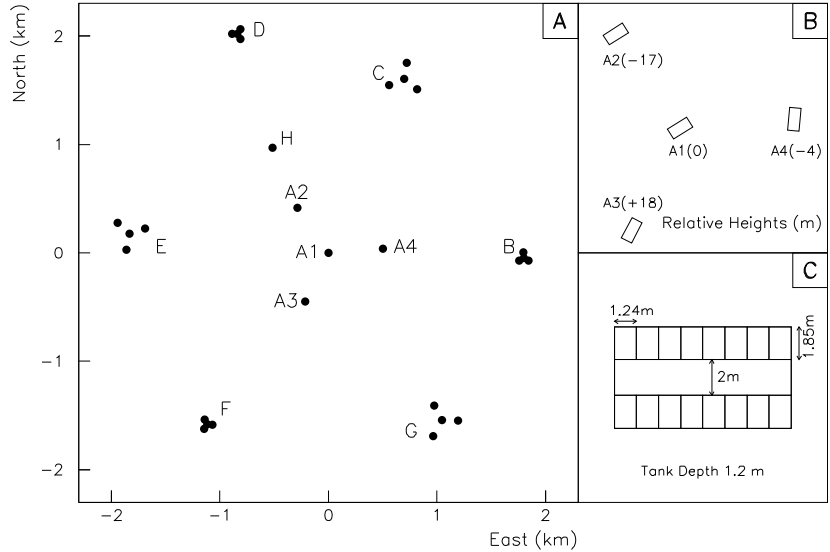}
\caption{The Haverah Park Array. A) The 2 km array. B) The relative 
heights and orientations of the A-site detector huts. C) The 
arrangement of water tanks within an A-site detector hut.}
\label{array.fig}
\end{figure}

The trigger rate of an air-shower array at large zenith angles is 
extremely sensitive to the geometry of the array. Factors such as 
the shape and relative height of detectors become very important 
for such showers. Figure~\ref{array.fig}A shows the layout of the 
Haverah Park array. The relative heights and orientations of the 
four A-site detectors, the triggering detectors, are shown in 
figure~\ref{array.fig}B. A gradient across the array is apparent 
and this has a significant effect on the observed azimuthal 
distribution (see section~\ref{results}). Figure~\ref{array.fig}C
shows the positions of individual tanks within a detector hut. The 
signals from 15 of the 16 tanks, each of area 2.25~m$^{2}$, were 
summed to provide the signal used in the trigger. One tank in each 
hut was used to provide a low gain signal. See \cite{haverah} 
for a more detailed description of the array.

Water-\v Cerenkov densities are expressed in terms of the mean signal 
from a vertical muon (1 vertical equivalent muon or VEM). It has 
been shown that this signal was equivalent to approximately 14 
photoelectrons (pe) \cite{thesis}.

The formation of a trigger was conditional on: 1. A density of 
$>$0.3~VEM m$^{-2}$ in the central detector (A1) 2. At least 2 of 
the 3 remaining A-site detectors recording a signal of 
$>$0.3~VEM m$^{-2}$.

The rates of the triggering detectors were monitored daily.  Over the
life of the experiment, after correction for atmospheric pressure
effects, the rates of the detectors were stable to better than 5\%.

\subsection{\v Cerenkov signals of vertical particles}

The calculation of the water-\v Cerenkov signal from horizontal 
showers is complex. It is informative to consider first the 
simpler case of vertical showers. A GEANT simulation of the 
propagation of vertical electrons, 
gammas, and muons through Haverah Park tanks has been performed 
(WTANK \cite{Wtank}). The wavelength dependences of the physical 
properties of the tank have been taken from \cite{Wtank}
but normalised to the following peak values which are the 
best estimates for the Haverah Park tanks:
\begin{itemize}
\item reflectivity of the walls - 83\%
\item absorption length of the water - 15 m
\item photomultiplier (PMT) quantum efficiency - 22\%
\end{itemize}
The Thorn/EMI, 9618 PMT wavelength acceptance function is taken from 
the manufacturers specifications.

\begin{figure}
\ybox{0.35}{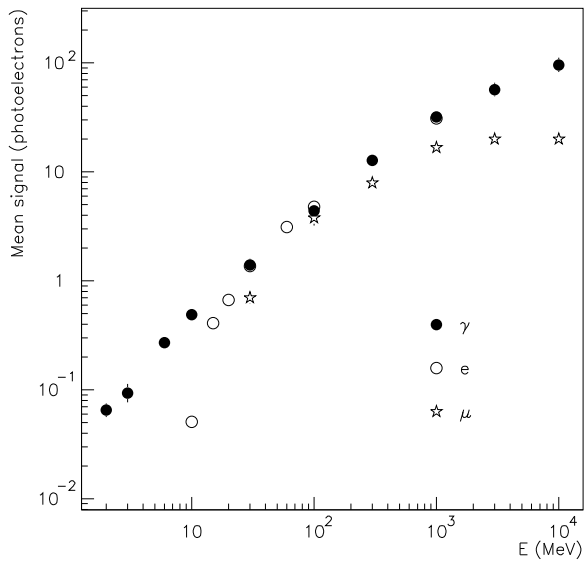}
\caption{The mean signal produced in a Haverah Park tank by 
vertical electrons, muons, and gammas as a function of energy 
(low energy electrons are absorbed in the lid of the tank).}
\label{tank_egm.fig}
\end{figure}

The results of this simulation are summarized in 
figure~\ref{tank_egm.fig}. It can be seen that the values used 
enable us to reproduce the measured signal from an average vertical 
muon (assuming a mean muon energy of 1~GeV \cite{Auger}). 

The signal from a vertical muon is composed primarily of the 
\v Cerenkov light emitted from the muon track. However there is a
significant contribution from \v Cerenkov light emitted by
$\delta$-ray electrons (2 pe of the 14 pe total for an average
vertical muon).

By contrast the mean energy of electrons and gamma-rays in vertical
showers is below 10 MeV. Convolving the energy spectrum of electrons
and gammas with the response given by figure~\ref{tank_egm.fig}, we
find mean signals of 2.6 and 0.9 photoelectrons for electrons and
gamma-rays respectively.

\subsection{Signals produced by horizontal showers}

Three factors complicate the picture for horizontal showers.  Firstly
for horizontal showers it is possible for \v Cerenkov photons to fall
directly onto the PMT without reflection from the tank walls (we refer
to such photons as ``direct light'').  Secondly the
azimuthal asymmetry of the Haverah Park tanks (and detector huts)
becomes increasingly important at large zenith angles. Thirdly the
mean muon energy grows with zenith angle.  For 87$^{\circ}$ showers
the mean muon energy is 200~GeV. For higher energy muons the
probability of interaction in the tank is much greater. The production
of secondary electrons via pair-production, bremsstrahlung, nuclear
interactions (collectively referred to as PBN interactions), and
electron knock-on ($\delta$-rays) is therefore enhanced. For example
the correction due to $\delta$-ray production increases from 2 pe at
typical vertical muon energies of 1 GeV to around 3 pe for $>10$ GeV.

\begin{figure}
\ybox{0.5}{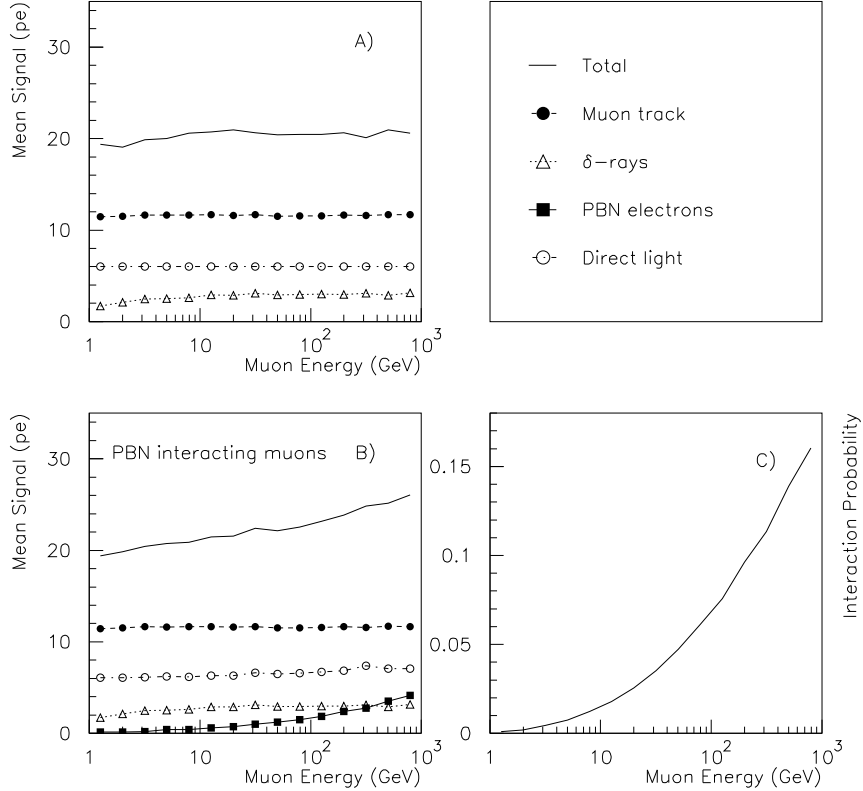}
\caption{The different contributions to the total signal produced
by horizontal muons. A) The mean signal produced by those muons not
undergoing PBN interactions in the tank. B) The mean signal produced
by PBN interacting muons C) The probability of a muon undergoing a
PBN interaction as a function of energy.} 
\label{tank_mv.fig}
\end{figure}

Signal enhancements due to direct light and muon interactions appear
in only a fraction of events and hence do not greatly affect the peak
of the photoelectron distribution. They do however add long tails to
this distribution which are of great importance in calculation of the
array trigger rate. Figure~\ref{tank_mv.fig} shows the different
contributions to the water-\v Cerenkov signal of a horizontal muon as
a function of energy. A single Haverah Park tank is considered with
muons propagating parallel to the short (1.24~m) side of the tank.

For showers of $>60^{\circ}$ the electromagnetic component is almost
entirely a product of muon decay as explained in section
\ref{emag}. Convolving the number and energy spectrum of electrons and
gammas with the tank response shown in figure \ref{tank_egm.fig} we
find that the electromagnetic component contributes 2.8 pe for each
muon corresponding to 20$\pm$2 \% of the vertical equivalent muon
signal.

\section{Implementation}
\label{method}

We have simulated the rate of the Haverah Park array as a function of
the zenith angle following
the steps described below. First of all we simulated 100, 10$^{19}$~eV,
showers for zenith angles from $60^{\circ}$ to $88^{\circ}$ in steps
of $2^{\circ}$, without magnetic field, for each of the four
combinations of primary mass and interaction model as described
above. Use is made of the ideas developed in Paper I.

For each zenith angle the following procedure is applied:
\begin{enumerate}
\item The inputs of the analytical model (LDF, ${\cal E} (r)$ and $<d>$) 
are parameterized.
\item The analytical model is applied to these 
parameterizations to generate muon density maps of the showers in 
the transverse plane taking into account magnetic field effects.
\item The density maps for different primary energies  are obtained 
with the energy scaling relationship given by Eq.~\ref{Escaling} 
and the parameters shown in Table \ref{nmu.tab}.
\item Different azimuthal angles are considered by simple 
rotation algorithms in the approximation explained in section~3 of 
Paper I.
\end{enumerate}

For each of these zenith angles we have 
calculated the triggering probability of the Haverah Park array 
in 40 energy bins between $10^{16}$~eV and $10^{20}~$eV and 18 
azimuth bins as follows:

\begin{enumerate}
\item The shower is directed on to the array N times with random 
core locations on the ground plane up to X km from the centre of 
the array. N and X are varied according to the primary energy 
and zenith we are dealing with. Typical values are $N=10000$ and 
$X=8$~km. 

\item Each time a shower is directed at the array, the total muon 
number ($N_{\mu}$) is fluctuated with a gaussian 
distribution of spread 0.2$N_{\mu}$ to take into 
account shower fluctuations.
\item The density in the ground plane at the location of each 
of the trigger detectors is read from the muon density maps with 
an appropriate projection, taking into account the corrections due 
to the different heights of the detectors and the effects of 
the magnetic field as explained in section~7 of Paper I. 
\item The corresponding signal in each of the trigger detectors 
is generated (see next subsection).
\item The trigger condition of the Haverah Park array is tested.
\item If an event is deemed to trigger the array then the 
primary energy and the combination of detectors contributing 
to the trigger are recorded.  
\end{enumerate}

\begin{figure}
\ybox{0.4}{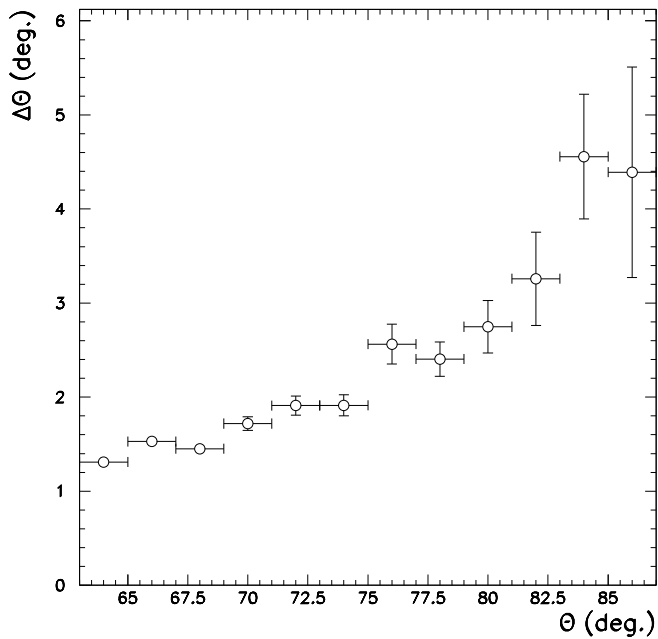}
\caption{The uncertainty in reconstructed zenith angle as a function of zenith angle.} 
\label{theta_err.fig}
\end{figure}

The triggering probability derived in this way has been convolved with
a primary energy spectrum \cite{WatsonNagano}, derived from Akeno and
Haverah Park data, to obtain the event rate at each zenith and azimuth
angle. The final step before comparison with data is to convolve the
obtained zenith angle distribution with the appropriate measurement
errors.  The arrival directions of all the data that had zenith angle
greater than $60^\circ$ have been reanalysed assuming a plane front for 
the muons and using all available timing information. Previously only 
times from the four central triggering detectors were used to compute 
the arrival directions \cite{zenith1st}.  Here up to 16 times were 
used in the analysis with the additional data being from the detectors
$\sim$2~km from A1 in the ground plane.
The uncertainly in arrival 
direction for each event has accordingly been significantly reduced.  
Fig. \ref{theta_err.fig} shows the average 
uncertainty in zenith angle of the reconstructed events as a function of 
zenith angle. The uncertainty in azimuth angle is approximately constant 
and is $\sim 1^{\circ}$. The events considered here have not been
analysed for core position. Accordingly some of the very large zenith
angle events may have fallen a long way outside the boundary of the
array and consequently have a larger uncertainty in zenith angle 
than suggested by fig. \ref{theta_err.fig}.

We have rejected about $3\%$ of the events having an error exceeding
$4.5^{\circ}$ in the reconstructed zenith angle $\theta$.  The RMS
error in zenith angle shown in fig. \ref{theta_err.fig} is used to \emph{smear} 
the calculated distribution.

\subsection{Implementation of the signal in detector}

The signal is generated as follows:

\begin{enumerate}
\item The projected area of the detector in the shower plane is calculated.
\item Given the local muon density and the projected area, we sample the number of 
      incident muons from a Poisson distribution.
\item The track length of each muon through the detector is sampled from a
      distribution obtained analytically from the detector geometry
      (see figure~\ref{array.fig}C).\footnote{This distribution accounts for the 
      possibility that a single muon may traverse several tanks.} 
\item The contribution of indirect \v Cerenkov light from the incident muons 
      and from $\delta$-ray electrons is calculated from the sampled track
      lengths (12 pe for each 1.2 m of track, with an additional 3 pe/1.2 m
      to account for the signal from $\delta$-rays as described in the 
      previous section).
\item The signal from direct light on the PMTs is related to the detector geometry
      in a more complex way and is implemented using WTANK to simulate the
      passage of muons through the whole detector for a range of zenith and
      azimuth angles.
\item The probability of PBN interactions in the detector is calculated
      from the track length and the muon energy (obtained from the ${\cal E} (r)$ 
      with a zenith angle dependent correction for the magnetic field).
\item If an interaction is deemed to occur then the energy of the resulting 
      em-cascade is sampled from the appropriate differential cross-section.  
      The signal produced by this cascade is calculated using an analytical approach.
\item The electromagnetic component of the shower is approximated by the addition 
      of 2.8 pe for each muon as discussed in the previous section. 
\end{enumerate}

The contributions to the signal depend in different ways on the zenith
and azimuth.  The mean signal for those contributions proportional to
the track are constant with azimuth and zenith, because of the
compensating effect of the projected area of the detector. In fact the
mean signal is simply proportional to the volume of the detector. For
larger zenith angles fewer muons produce the same signal so Poisson
fluctuations become more important.  The mean signal produced by direct light
increases with zenith and has a complex dependence on azimuth. The mean signal
 from the electromagnetic component is proportional to the projected area, and 
hence decreases with zenith and has a simpler azimuthal dependence.  

We note that the use of distributions (rather than 
mean values) for different contributions to the signal is essential 
for accurate calculation of the event rate.

\section{Results and comparison with data}
\label{results}

\subsection{Energy distribution of showers that trigger the array}

\begin{figure}
\ybox{0.5}{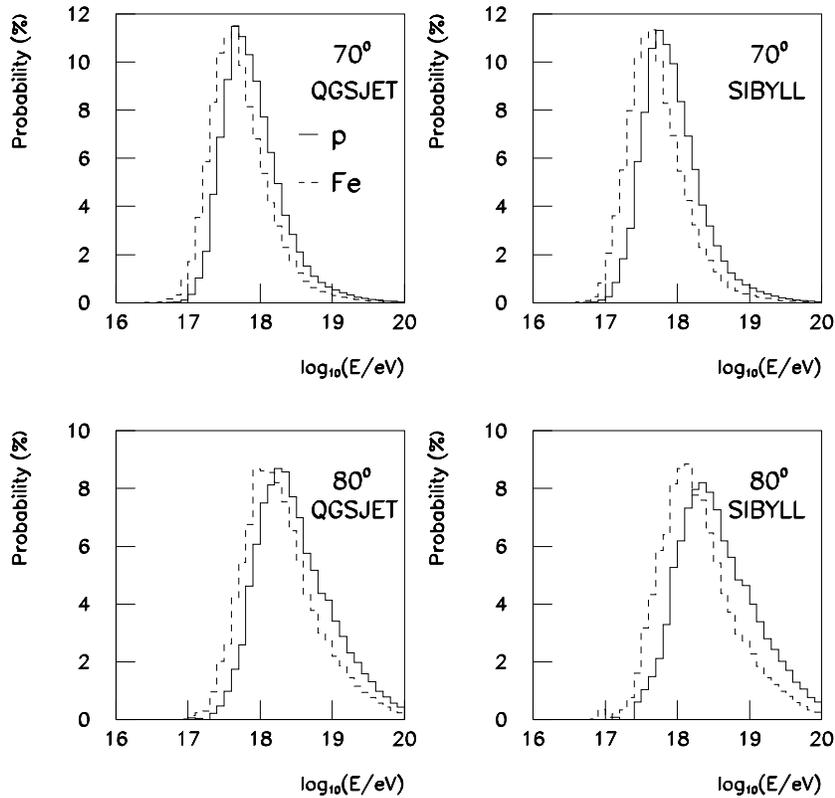}
\caption{The energy distribution of showers which trigger the Haverah Park array at 70$^\circ$
and 80$^\circ$ for each of the four combinations of model and 
primary mass composition shown as a probability for each energy bin.} 
\label{ene_response.fig}
\end{figure}

In Fig.~\ref{ene_response.fig} we show the energy distribution of showers
which trigger the array as calculated with
the QGSJET and SIBYLL models and for proton and iron primaries
at two zenith angles. The mass dependence seen in Fig.~\ref{ene_response.fig}
is a consequence of the greater number of muons produced by heavier primaries.
The choice of interaction model has a some what smaller effect on the energy 
response. 

\begin{figure}
\ybox{0.5}{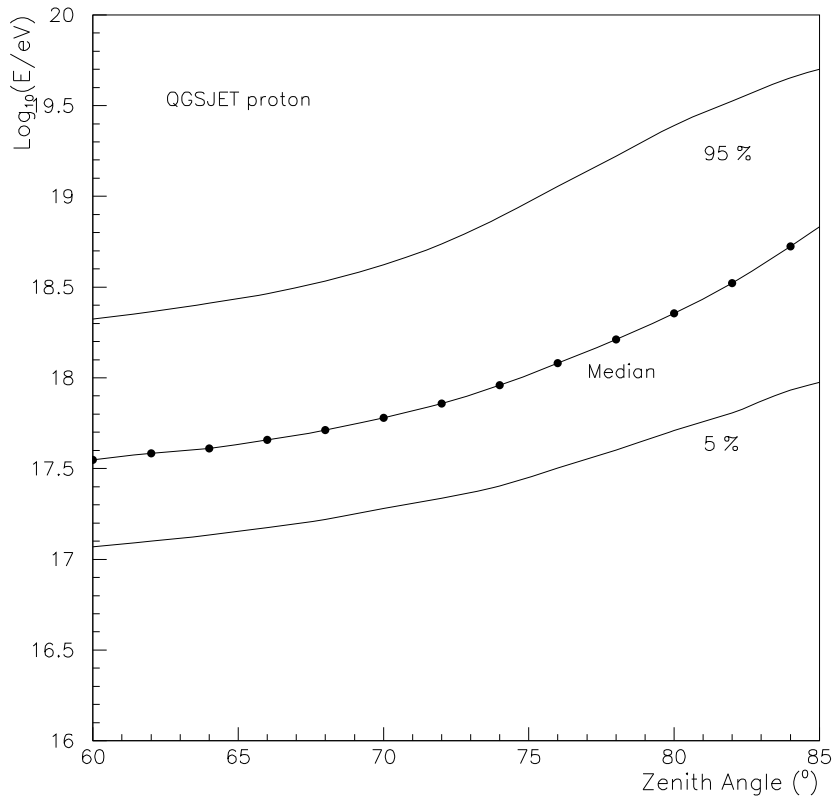}
\caption{The median and percentiles of the energy distributions of 
Fig.~\ref{ene_response.fig} as a function of zenith angle (QGSJET with proton
proton primaries).} 
\label{median.fig}
\end{figure}

Fig.~\ref{median.fig} shows the mean and width of the energy distribution 
as a function of zenith angle (for QGSJET with proton primaries).
The median energy changes from 
$\sim 6\times 10^{17} $ eV at 70$^\circ$ to 
$\sim 5\times 10^{18} $ eV at 84$^\circ$.

\subsection{Azimuthal distribution}

\begin{figure}
\ybox{0.5}{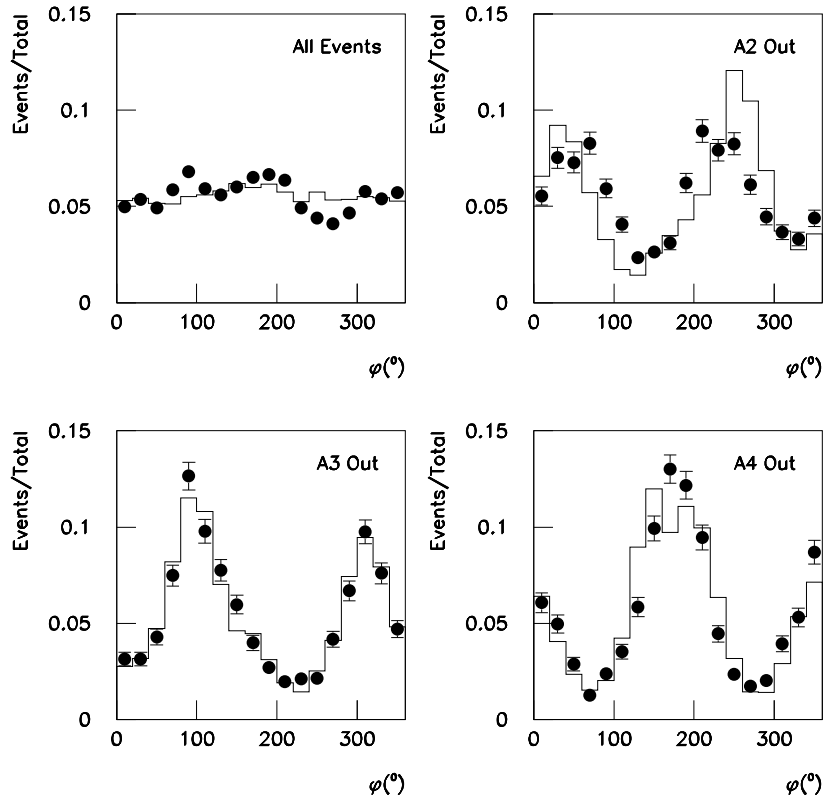}
\caption{Azimuthal distribution of events recorded above 59$^\circ$ 
for all events and three sets of three fold events. The distribution
of events not triggering each of the A2, A3 and A4 detectors are 
shown in the later three plots. The simulations use the QGSJET model
with proton primaries normalised to the data.}
\label{azy.fig}
\end{figure}

One of the most stringent tests for the simulation of horizontal air
showers is the prediction of the azimuthal distributions. Minor
corrections in the altitude of the detectors, or the orientation of
the tanks can lead to large differences in the rates for inclined
showers. The total event rate is dominated by events triggering only 3
of the 4 A-site detectors. The azimuthal distribution is best
understood by separating these 3-fold events into 3 sets, each set
requiring that a particular detector \emph{does not} trigger. The
recorded azimuthal distributions for each of these three sets (A2 out,
A3 out and A4 out) are shown in Fig.~\ref{azy.fig}. The calculated
distributions (using QGSJET with primary protons) are shown 
normalised to the data for comparison of the shape.
The agreement is reasonable with the exception of the set ``A2 out''. 
There are many possible reasons for this disagreement. In particular 
we will in future consider reflection and absorption in the ground 
around the detectors, but we are encouraged by the overall result.

\subsection{Zenith angle distribution}

\begin{figure}
\ybox{0.5}{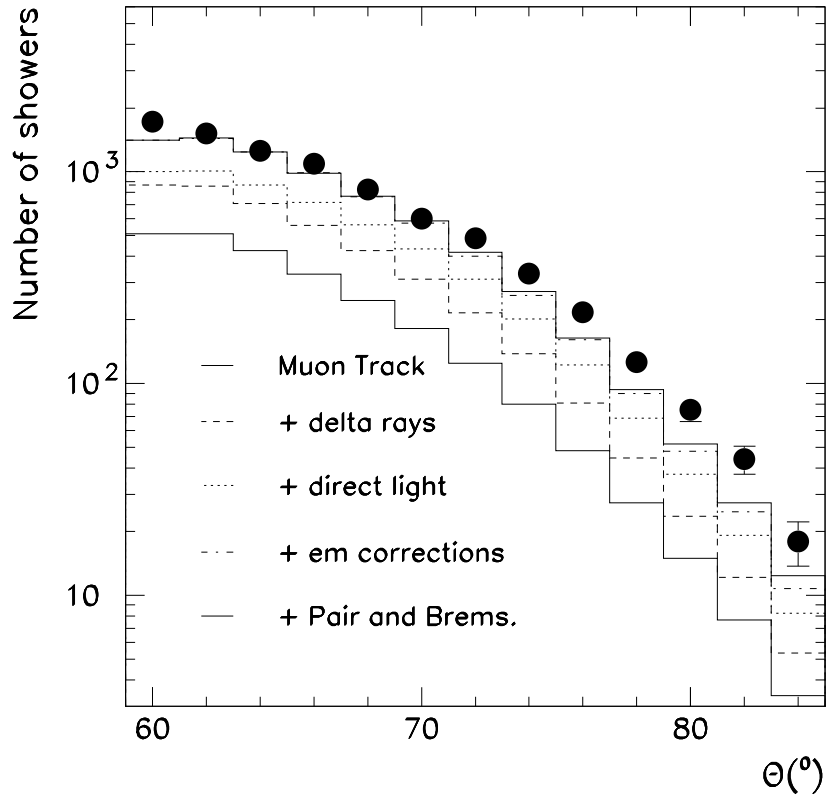}
\caption{The total number of events recorded by the Haverah Park 
array as a function of zenith (dots) compared with the prediction
obtained using a proton primary composition and the QGSJET model.
Different contributions to the total rate are illustrated and are added 
cumulatively.} 
\label{rate_steps.fig}
\end{figure}

In Fig.~\ref{rate_steps.fig} we show the contributions to the total
rate (from each of the processes considered) as a function of zenith
angle. The QGSJET model and proton primaries are assumed. Similar
plots for other hadronic models and primaries differ mainly in their
normalization. It should be noted that inclusion of only the muon
track contribution would lead to underestimating the total rate by a
factor $\sim 3$ ($\sim 4$) at 65$^\circ$ (85$^\circ$).  The effects of
delta rays, direct light, electromagnetic corrections, and pair
production and bremsstrahlung are all very significant in the
calculation of the overall rate.  The effect of fluctuations in the
total muon number in a shower represents at most a $20 \%$ rate
enhancement because of the large width of the energy response at a
given angle.

\begin{figure}
\ybox{0.32}{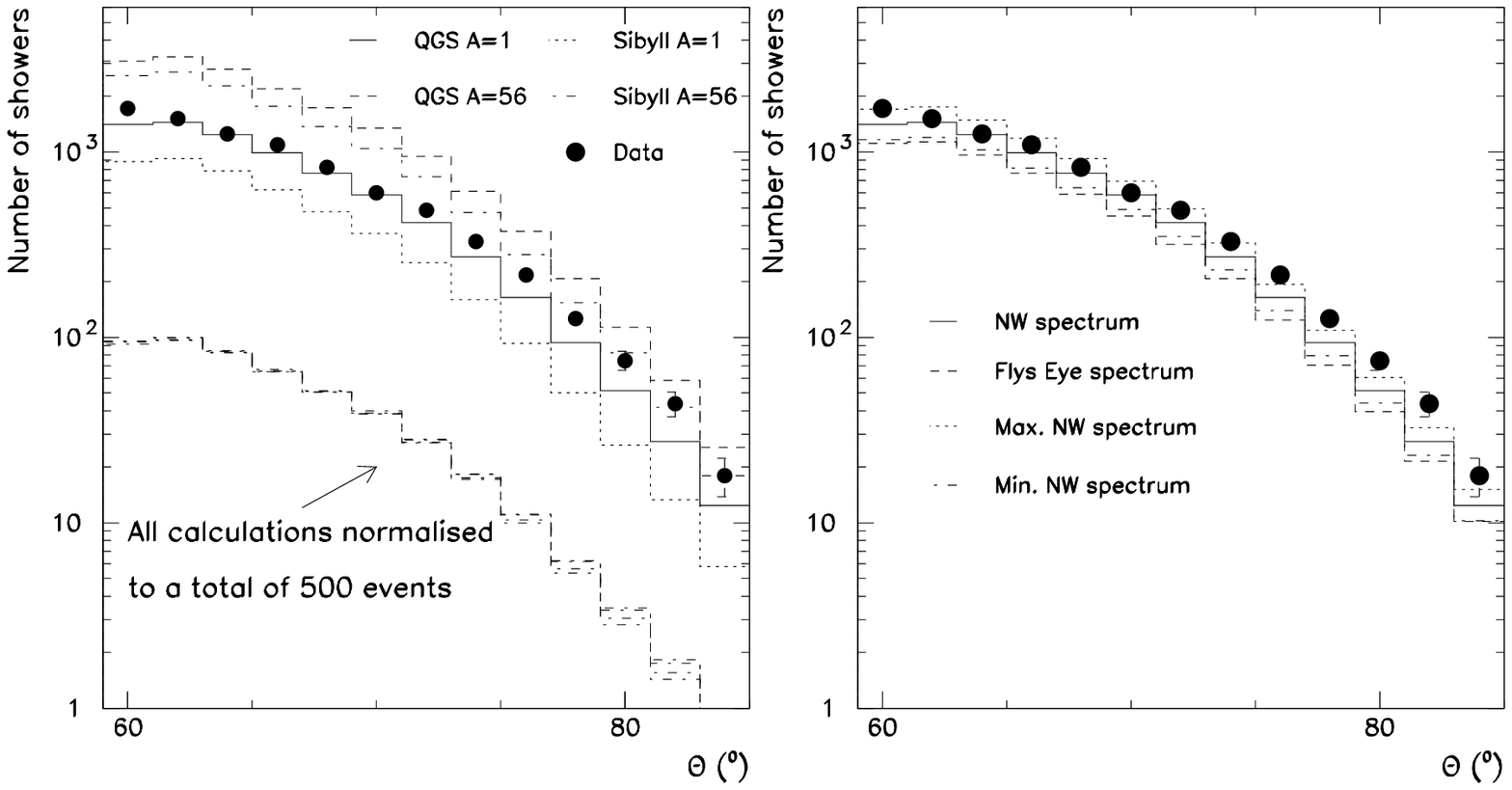}
\caption{Left panel: A comparison of all four model/mass 
combinations at the recorded Haverah Park zenith angle distribution.
The lower histograms show the calculated rates normalized to 
illustrate the differences in shape. Right panel: A comparison
of the recorded Haverah Park zenith angle distribution for
different primary spectra.}
\label{rate.fig}
\end{figure}

The left hand plot of Fig.~\ref{rate.fig} shows the number of events
detected at Haverah Park as a function of the zenith angle compared to
the expected number of events for two different models and two mass
compositions.  The error bars on the data points are statistical.  The
right hand plot illustrates the effect of the uncertainty
in the primary energy spectrum on our result. The maximum and minimum
predictions from the spectrum described in \cite{WatsonNagano} (using
one sigma errors) are shown together with the result obtained using
the primary spectrum parameterized in \cite{flyeseye} (all for proton
primaries using the QGSJET model).

We stress that no normalization was done with the simulated events. 
The shape of the simulated distribution is in good
agreement with data between $60^{\circ}$ and $70^{\circ}$ while at 
higher zenith angles there is a mild disagreement, at most a 
$\sim $30\% effect. 
Within the uncertainties resulting from the zenith angle measurement, the primary mass, 
the interaction models
and the primary energy spectrum, the simulated overall rate 
of showers above $60^{\circ}$ is completely consistent with the data. 

For most assumptions of primary spectrum and interaction model an
intermediate primary mass in the decade around $10^{18}$~eV seems 
to be required to fit the data. 
Further work is in progress to select higher energy events 
by considering detectors (B -- H of Fig.~\ref{array.fig}A) that are not in 
the trigger. For example the event discussed in \cite{Hillas} and believed to 
be of energy $\sim 10^{20}~$eV struck 20 of the detectors.  

\section{Conclusion}
\label{conclusions}

We have presented the first comparison of $>10^{17}$~eV air-shower 
data above $60^{\circ}$ with simulation. The agreement is reasonable
and allows us to state that the detection of HAS induced by 
hadronic primaries in an array of water-\v Cerenkov tanks can be 
explained in terms of the muonic component. The effects 
of all the considered contributions to the tank signals 
have been shown to be extremely important in understanding the 
total rate.

Our final results differ from the preliminary ones reported earlier
\cite{utah99} in several respects. We have taken into consideration
the electromagnetic component from muon decay separately.  Previously
this contribution was taken together with the correction due to
$\delta-$rays. We have further improved the implementation of the PBN
corrections taking into account their increase with the muon energy.
We have also simulated the ground number density profiles in a
completely different fashion making use of the model described in
Paper I.

The general shape agreement between data and simulation both for the
zenith angle and azimuthal angle differential rates, together with
agreement in the absolute normalization makes us confident that the
geometrical considerations have been taken into account correctly, at
least at the 20\% level.  This, together with the general agreement
with our preliminary rate simulation, which used Monte Carlo shower
simulation rather than the analytical approach described in Paper I,
provides strong supporting evidence for this approach as an important
tool for HAS studies.

The normalization of the shower rate is sensitive to the model used
for high energy hadronic interactions. For proton showers the absolute
normalization of the rate differs by up to a factor of two for the two
models considered. The absolute rate has also been shown to be very
sensitive to the primary composition. There is about a factor 2.2 (3)
enhancement, reasonably independent of zenith angle, for the
differential rate normalization if the primary particles are assumed
to be iron nuclei in the QGSJET (SIBYLL) model.  It is interesting 
to note that current air-shower models are not able to reproduce
the observations of high energy particles close to the shower axis
above 10$^{16}$~eV \cite{kascade}. This may be a possible explanation of
the $\sim$30\% disagreement above 80 degrees (where the surviving muons
all originate close to the shower core). 

These effects illustrate that HAS measurements may provide another
tool with which to study both primary composition and hadronic
interactions in ultra high energy cosmic rays. Future detectors such
as the Auger Observatories should benefit significantly from the study
of cosmic ray showers at large zenith angles.

{\bf Acknowledgements:} We thank Xavier Bertou for helping us with the
angular reanalysis of the data described in \cite{utah99} and Gonzalo
Parente for suggestions after reading the manuscript. This work was
partly supported by a joint grant from the British Council and the
Spanish Ministry of Education (HB1997-0175), by Xunta de Galicia
(XUGA-20604A98), by CICYT (AEN99-0589-C02-02) and by PPARC(GR/L40892).
\end{document}